\documentclass[aps,pre,reprint,groupedaddress]{revtex4-1}
\usepackage{amsmath}

\usepackage{bbding} 
\usepackage{amssymb}
\usepackage{amsfonts}
\usepackage{dcolumn}
\usepackage[framemethod=TikZ]{mdframed}
\usepackage{xcolor}
\usepackage{setspace}
\usepackage{caption}
\usepackage{subcaption}
\usepackage{lipsum}
\usepackage{tcolorbox} 
\usepackage{mathtools} 
\usepackage{tgpagella}

\newcommand{\refew}[1]{Eq.\eqref{eq:#1}}
\newcommand{\reffig}[1]{Figure \ref{fig:#1}}

\newcommand{\mm}[0]{\nonumber \\}
\newcommand{\tbf}[1]{\textbf{#1}}
\newcommand{\lbar}[0]{{\overline{\lambda}_0}}

\begin{document}


\title{Statistical Physics of the Symmetric Group }



\author{Mobolaji Williams} 
\affiliation{%
Department of Physics, Harvard University, Cambridge, MA 02138, USA 
}%
\email{mwilliams@physics.harvard.edu}

 \author{Eugene Shakhnovich}%
\affiliation{%
Department of Chemistry and Chemical Biology, Harvard University, Cambridge, MA 02138, USA 
}%
 \email{shakhnovich@chemistry.harvard.edu }


\date{\today}

\begin{abstract}

Ordered chains (such as chains of amino acids) are ubiquitous in biological cells, and these chains perform specific functions contingent on the sequence of their components. Using the existence and general properties of such sequences as a theoretical motivation, we study the statistical physics of systems whose state space is defined by the possible permutations of an ordered list, i.e., the symmetric group, and whose energy is a function of how certain permutations deviate from some chosen correct ordering. Such a non-factorizable state space is quite different from the state spaces typically considered in statistical physics systems and consequently has novel behavior in systems with interacting and even non-interacting Hamiltonians. Various parameter choices of a mean-field model reveal the system to contain five different physical regimes defined by two transition temperatures, a triple point, and a quadruple point. Finally, we conclude by discussing how the general analysis can be extended to state spaces with more complex combinatorial properties and to other standard questions of statistical mechanics models.

\end{abstract}


\maketitle


\section{Introduction \label{sec:one}}

Chains of amino acids are important components of biological cells, and for such chains the specific ordering of the amino acids is often so fundamental to the resulting function and stability of the folded chain that if major deviations from the correct ordering were to occur, the final chain could fail to perform its requisite function within the cell, proving fatal to the organism. 

\begin{figure}[h]
\centering
\includegraphics[width=\linewidth]{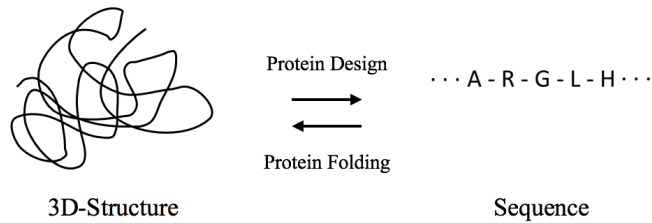}
	\caption{Folding vs. Design (or Inverse Folding) problems: The protein folding problem is concerned with determining the three dimensional structure produced by a particular sequence of amino acids. The protein design problem (which motivates the current work) is concerned with finding the sequence(s) of amino acids which yield a given three dimensional polypeptide structure. A number of approaches to the design problem are given in \cite{shakhnovich1998protein}}
\label{fig:design}
\end{figure}

More specifically, we see the relevance of correct ordering in the study of protein structure, which is often divided into the protein folding and protein design problem. While the protein \textit{folding} problem concerns finding the three-dimensional structure associated with a given amino acid sequence, the protein \textit{design} problem (also termed the inverse-folding problem; see \reffig{design}) concerns finding the correct amino acid sequence associated with a given protein structure. \\
\indent An aspect of one solution to the protein design problem is to maximize the energy difference between the low-energy folded native structure and the higher energy misfolded/denatured structures. In doing so, one takes native structure as fixed and then determines the sequence yielding the minimum energy, under the assumption (termed the "fixed amino-acid composition" assumption) that only certain quantities of amino-acids appear in the chain  \cite{morrissey1996design}. In this resolution (specifically termed heteropolymer models \cite{shakhnovich1993engineering} \cite{shakhnovich1993new}) the correct amino acid sequence is found by implementing an MC algorithm in sequence space given a certain fixed amino acid composition. This entails assuming the number of various types of amino acids does not change, and distinct states in sequence space are permutations of one another. For example, for a polypeptide chain with $N$ residues, rather than searching over the entire sequence space (of size $20^N$), one searches over a space of sequences (of size $N!/n_1! n_2!\ldots n_{20}!$) which are defined by a fixed number of each amino acid.

This aspect of the protein design problem alerts one to a gap in the statistical mechanics literature. Namely, there do not seem to be any simple and analytically soluble statistical mechanics models where the space of states is defined by permutations of a list of components.

We can take steps toward constructing such a model by considering reasonable general properties it should have. If we assume there was a specific sequence of components which defined the lowest energy sequence and was thermodynamically stable in the model, then deviations from this sequence would be less stable. Because of the role sequences of molecules play in biological systems, it is worth asking what features we expect such sequences to have from the perspective of modeling in statistical mechanics.

In Section II we introduce the model, and compute an exact partition function which displays what we term ``quasi"-phase transitions---a transition in which the sequence of lowest energy becomes entropically disfavored above a certain temperature. In Section III, we extend the previous model by adding a quadratic mean field interaction term and show that the resulting system displays two transition temperatures, a triple point, and a quadruple point. In Section IV, we discuss various ways we can extend this model in theoretical or more phenomenological directions.

\section{System and Partition Function \label{sec:two}}
Our larger goal is to study equilibrium thermodynamics for a system defined by permutations of a set of $N$ components where each unique permutation is defined by a specific energy. In general, we should consider the case where the set of $N$ components consists of $L$ types of components for which if $n_k$ is the number of repeated components of type $k$, then $\sum_{k=1}^{L} n_{k} = N$. For simplicity, however, we will take $n_k =1$ for all $k$ so that each component is of a unique type and $L = N$. 

To study the equilibrium thermodynamics of such a system with a fixed $N$ at a fixed temperature $T$, we need to compute its partition function. For example, for a sequence with $N$ components (with no components repeated), there are $N!$ microstates the system can occupy and assuming we label each state $k=1, \ldots , N!-1, N!$, and associate an energy $\epsilon_{k}$ with each state, then the partition function would be 
\begin{equation}
Z = \sum_{k=1}^{N!} e^{-\beta\epsilon_k},
\label{eq:correct}
\end{equation}
where $\epsilon_k$ for each state $k$ could be reasoned from a more precise microscopic theory of how the components interact with one another. Phenomenologically, \refew{correct} would be the most precise way to construct a model to study the equilibrium properties of permutations, but because it bears no clear mathematical structure, it is unenlightening from a theoretical perspective. \\ \indent Instead we will postulate a less precise, but theoretically more interesting model. For most ordered chains in biological cells, there is a single sequence of components which is the ``correct" sequence for a particular macrostructure. Deviations from this correct sequence are often disfavored because they form less stable macrostructures or they fail to perform the original function of the ``correct" sequence. With the general properties of such sequences in mind, we will abstractly represent our system as consisting of $N$ sites which are filled with particular coordinate values denoted by $\omega_k$. That is, we have an arbitrary but fixed coordinate vector $\vec{\omega}$ expressed in component form as 
\begin{equation}
\vec{\omega} = (\omega_1, \ldots, \omega_N).
\end{equation}
We will take the collection of components $\{\omega_k\}$ as intrinsic to our system, and thus take the state space of our system to be the set of all the vectors whose ordering of components can be obtained by permuting the components of $\vec{\omega}$, i.e., all permutations of $\omega_1, \ldots, \omega_N$. We represent an arbitrary state in this state space as $\vec{\theta} = (\theta_1, \ldots, \theta_N)$, where the $\theta_k$ are drawn without repeat from $\{\omega_k\}$.  Formally, we would say our space of states is isomorphic to the symmetric group on $\vec{\omega}$ (\cite{dixon1996permutation}). We will thus denote our state space as 
\begin{equation}
Sym(\omega) \,\,:=\,\, \text{Set of All Permutations of $(\omega_1, \ldots, \omega_N)$}.
\end{equation}
and then an arbitrary state $\vec{\theta}$ is just an element element of this set. \\
\indent As a first formulation of the model, we will take $\vec{\theta}_0 = \vec{\omega}$ (the correct sequence) to represent the zero energy state in the system, and for each component $\theta_i$ of an arbitrary vector $\vec{\theta}$ which differs from the corresponding component $\omega_i$ in $\vec{\omega}$, there is an energy cost of $\lambda_i>0$. The Hamiltonian is then
\begin{equation}
{\cal H}_{N}(\{\theta_{i}\}) =  \sum_{i=1}^N \lambda_i \,I_{\theta_{i} \neq \omega_{i}},
\label{eq:haminit01}
\end{equation}
where $\theta_{i}$ and $\omega_{i}$ are components of vectors $\vec{\theta}$ and $\vec{\omega}$ respectively, and $I$ is defined by 
\begin{equation}
I_{A} \equiv \begin{dcases} 1  & \text{if $A$ is true} \\ 0 & \text{if $A$ is false} \end{dcases}.
\end{equation}
We note that although we label our general state as $\vec{\theta} = (\theta_1, \ldots, \theta_N)$, the components $\theta_1, \ldots, \theta_N$ can only take on mutually-exclusive values from the set $\{\omega_{k}\}$.\\
\indent We want to explore the equilibrium thermodynamics of a system with a Hamiltonian of \refew{haminit01}. This amounts to calculating the partition function 
\begin{equation}
Z_{N}(\{\beta\lambda_i\}) = \sum_{\vec{\theta} \,\in\, Sym(\omega)} \text{exp}\left(- \beta \sum_{i=1}^N\lambda_i\, I_{\theta_{i} \neq \omega_{i}}\right),
\label{eq:partfunc01}
\end{equation}
where $Sym(\omega)$ is again the set of all permutations of the components of $(\omega_1, \ldots, \omega_N)$. To find a closed form expression for the partition function, we group terms in \refew{partfunc01} according to the number of ways to completely reorder $j$ components in $\vec{\omega}$ while keeping the remaining components fixed. Each such reordering (i.e., each value of $j$) is associated with a sum over products of $e^{-\beta\lambda_i}$ terms with $j$ factors of $e^{-\beta\lambda_i}$ (for various $i$) in each term.  The total partition function is a sum of all such reorderings for all $j$s from $0$ to $N$. As can be seen from a direct expansion of \refew{partfunc01}, we have
\begin{equation}
Z_{N}(\{\beta\lambda_i\}) = \sum_{j =0}^N d_{j} \,\Pi_j\left(e^{-\beta\lambda_1}, \ldots, e^{-\beta\lambda_N}\right),
\end{equation}
where $d_j$, termed the number of derangements of a list of $j$ (\cite{chuan1992principles}), is the number of ways to completely reorder a list of $j$ elements. The quantity $\Pi_{j}(x_1, \ldots, x_N)$, termed the $j$th elementary symmetric polynomial on $n$ (\cite{borwein2012polynomials}), is the sum of all ways to multiply $j$ elements out of the $N$ term set $\{x_1, \ldots, x_N\}$. For example, $\Pi_{2}(x_1,x_2, x_3) = x_1 x_2 + x_2 x_3 + x_3 x_1$. By definition 
\begin{equation}
\Pi_{k}(x_1, \ldots, x_N) = \frac{1}{k!} \left[ \frac{d^k}{d q^k} \prod_{i = 1}^{N} (1+ q\, x_i) \right]_{q = 0}. 
\end{equation}
By the definition of the incomplete gamma function as  $\Gamma(x, \alpha)  = \int^{\infty}_{\alpha} dt\, t^{x-1} e^{-t} $ and its relation to derangements (i.e., $d_j = \Gamma(j+1, -1)/e$, see   \cite{weisstein2002incgamma}), we then find 
\begin{align}
Z_{N}(\{\beta\lambda_i\}) &  = e^{-1}\int^{\infty}_{-1} dt\, e^{-t} \sum_{j=0}^N t^j\, \Pi_j\left(e^{-\beta\lambda_1}, \ldots, e^{-\beta\lambda_N}\right)\mm
& = \int^{\infty}_{0} ds\, e^{-s} \prod_{\ell =1}^{N} \Big[ 1 + (s-1) e^{-\beta\lambda_{\ell}}\Big],
\label{eq:partfunc001}
\end{align}
which is the desired closed-form expression for the partition function of this system. \\
\indent With \refew{partfunc001}, the problem of abstractly studying a thermal system of permutations with Hamiltonian \refew{haminit01} is, from the perspective of equilibrium statistical mechanics, now complete. However, there are still some physical and theoretical results which can be teased from this formalism. Specifically, we can ask whether this system exhibits phase transitions. To answer this question, it would prove more analytically tractable to take $\lambda_i = \lambda_0$ for all $i$. With this condition, our partition function simplifies to 
\begin{align}
Z_{N}(\beta\lambda_0) &  = \sum_{j=0}^{N} g_{N}(j) e^{-j \beta\lambda_0}  \label{eq:partfunc0} \\
& = \int^{\infty}_{0} ds\,  e^{-s}\left(1 + (s-1) e^{-\beta\lambda_0} \right)^N \label{eq:partfunc}
\end{align}
where we transformed our Hamiltonian as ${\cal H}_{N}(\{\theta_{i}\})  \to {\cal H}(j) = \lambda_0 j$ with $j$, defined as 
\begin{equation}
j \equiv \sum_{i=1}^{N}I_{\theta_i \neq \omega_i},
\label{eq:jdef}
\end{equation}
the number of components of $\vec{\theta}$ which are not equal to the corresponding component in $\vec{\omega}$. We call $j$ the number of \textit{incorrect} components of $\vec{\theta}$, and if $j=N$ we say $\vec{\theta}$ is \textit{completely disordered}. The factor $g_{N}(j)$ in \refew{partfunc0} is defined as 
\begin{equation}
g_{N}(j) = \binom{N}{j} d_j
\label{eq:gNj}
\end{equation}
is the number of ways to reorder a list of $N$ elements so that $j$ elements are no longer in their original position. This combinatorial definition of $g_{N}(j)$ will prove useful when we explore the phase behavior of more complex models of permutations.  

From the form of \refew{partfunc0}, it is clear that, physically, its associated Hamiltonian is not realistic as it places distinct permutations (which in any true physical system most likely have quite different energy properties) in the same degenerate energy state. Still, from a theoretical perspective, the simplicity of this model makes it a suitable starting point for studying the general properties of systems of permutations.  

\subsection{Phase-like Behavior of ``Non-Interacting" System}

We can investigate the phase-like behavior of the system defined by the Hamiltonian \refew{haminit01} (for constant $\lambda_i$ across $i$), by applying steepest descent to \refew{partfunc} in the $N\gg1$ limit. Doing so, we find we can approximate the free energy of the system to be 
\begin{align}
\beta F & =  - \ln Z_{N}(\beta\lambda_0) \mm
& \simeq N\beta\lambda_0 - \left(e^{\beta\lambda_0} - N-1\right) + F_0 (N),
\label{eq:free_en1}
\end{align}
where $F_0(N)$ is a $\beta\lambda_0$ independent constant. Noting that  $\langle j \rangle  = - \partial \ln Z_N(\beta\lambda_0)/\partial (\beta\lambda_0) = \partial F/\partial \lambda_0$, we find the average number of incorrect components satisfies the following equation of state:
\begin{equation}
\langle j \rangle \simeq N - e^{\beta\lambda_0}.
\label{eq:eos}
\end{equation}
By \refew{jdef}, we can infer that $\langle j\rangle$ must be greater than or equal to 0. However, the right-hand-side of \refew{eos} exhibits no such explicit constraint. Thus we can infer there is a phase-like transition in our system at the temperature
\begin{equation}
k_B T_c = \frac{\lambda_0}{\ln N}. 
\label{eq:crit1}
\end{equation}
Below this temperature, we must have $\langle j \rangle \simeq 0$ and thus the ``correct permutation" has the lowest free energy and is thermodynamically favored; above this temperature, $\langle j \rangle > 0$ and the system is in a disordered phase where the previous lowest energy ``correct-permutation" is energetically disfavored. \\
\indent Interestingly, this transition arises from the naively non-interacting Hamiltonian
\begin{equation}
{\cal H}_N (\{\theta_i\}) = \lambda_0 \sum_{i=1}^{N}I_{\theta_i \neq \omega_i}.
\label{eq:ham0}
\end{equation}
We say ``naively non-interacting" because \refew{ham0} consists of a sum over linear functions of a single index $i$, and thus does not suggest any coupling between terms of differing index.
However, statistical mechanics tells us that the energy of a system isn't the only thing which determines the thermodynamic behavior of a system. Indeed we have to consider entropic contributions as well, and in this system the entropy (as it is a function of $j$) can drive thermodynamic behavior. In other words, although the Hamiltonian is depicted as non-interacting and can set-theoretically be represented as
\begin{equation}
{\cal H}_{\text{system}} = {\cal H}_{1} \oplus {\cal H}_{2} \oplus \cdots \oplus {\cal H}_{N},
\end{equation}
our system really exhibits interactions between components because our total space of states ${\cal S}$ cannot be factorized:
\begin{equation}
{\cal S}_{\text{system}} \neq {\cal S}_{1} \otimes {\cal S}_{2} \otimes \ldots \otimes {\cal S}_{N}.
\end{equation}
Thus the ``non-interacting" system exhibits a transition at \refew{crit1} due to the coupled nature of the state space. As discussed in the subsequent section, we term this transition a ``quasi-phase transition" because it does not bear all of the standard properties we expect in phase transitions.

\subsubsection{Not a True Phase Transition}

\begin{figure}[t]
\centering
\includegraphics[width=.8\linewidth]{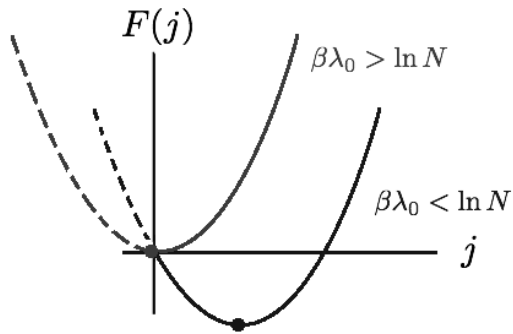}
	\caption{Free energy for ``Non-Interacting Model": For $\lambda_0>0$, the Landau  free energy of the system as a function of $j$, the number of incorrect components in $\vec{\theta}$, is always convex with a single global minimum. Because $j\geq0$, the $j<0$ domain of each plot (dashed section) is inaccessible. For sufficiently low temperatures, the minimum is at $j=0$, but as we increase the temperature beyond \refew{crit1}, the free energy curve moves to the right (but retains its functional form) and the new minimum is at a $j > 0$ value. Actual plots of \refew{freeen_0} for $j<0$ require us to replace the combinatorial argument of the logarithm with its corresponding gamma function expression.}
\label{fig:perm_nonint_freeenergy}
\end{figure}

We claim the system does not exhibit true phase transition behavior because many of these results are not consistent  with the traditional thermodynamic definition of phase transitions. For one, phase transitions are associated with divergences in the derivatives of the free energy, but there is no divergence in the free energy associated with the partition function \refew{partfunc} for possible parameter values. Also, the result \refew{eos} arises from the steepest descent approximation which makes $\langle j \rangle$'s temperature dependence near $\langle j\rangle=0$ appear non-differentiable when by \refew{partfunc} it is actually differentiable over its entire domain. Finally, with \refew{partfunc0} we can define a Landau free energy $F(j)$ for this system according to $Z = \sum_{j} e^{-\beta  F(j)}$, and what we may ordinarily label as a phase transition (i.e., going from $\langle j \rangle =0$ to $\langle j \rangle \neq 0$) arises, not from changes in the functional form of the Landau free energy as we see in real phase transitions, but from changes in the excluded region of the Landau free energy (See \reffig{perm_nonint_freeenergy}). Because the functional form of the free energy remains the same we observe no true phase transition.

Alternatively, a heuristic argument for the non-existence of phase transitions in our permutation model is mathematically very similar to the Landau argument (\cite{landau1980statistical}) for the non-existence of transitions in 1d Ising Models. For our permutation system with $N$ lattice sites, the state of zero energy and zero entropy consists of every site being occupied by its correct component. To increase the energy of this system, we can choose $j$ sites to contain incorrect components, thus giving us an energy ${\cal H}_{j} = \lambda_0 j$. The number of ways we can choose these $j$ components is given by \refew{gNj} Thus, upon introducing $j\neq 0$ incorrect components, the change in the Landau free energy of our system is 
\begin{align}
\Delta F(j)  
& =   \lambda_0 j - k_B T \ln \left[\binom{N}{j} d_{j}\right]  
\simeq j (\lambda_0  - k_BT  \ln N),  \label{eq:freeen_0}
\end{align}
where we took these results in the $1\ll j \ll N$ limit and used $d_j \simeq j!/e$. In the thermodynamic ($N\to \infty $) limit, we find that $\Delta F(j) \to - \infty$ meaning there is no non-zero $T$ at which the entropic contribution becomes subdominant to the energy. Thus the system exhibits no phase transition.

\section{Partition Function for Interacting Model \label{sec:three}}

When we first considered a model of thermal permutations, we began with a Hamiltonian where sites did not interact with one another and each had a site-dependent energy cost for being incorrectly occupied:
\begin{equation}
{\cal H} ( \{\theta_i\}) =\sum_{i} \lambda_i\, I_{\theta_i \neq \omega_i}.
\end{equation}
 More generally, we can consider Hamiltonians with an arbitrary number of multiple-site interaction terms. Such a Hamiltonian could be written as 
\begin{equation}
{\cal H} ( \{\theta_i\}) = \sum_{i} \lambda_i \, I_{\theta_i \neq \omega_i}  +\frac{1}{2}\sum_{i, j} \mu_{ij}\,I_{\theta_i \neq \omega_i} I_{\theta_i \neq \omega_i}+ \cdots.
\label{eq:haminit_gen0}
\end{equation}
The first term in \refew{haminit_gen0} associates an energy cost of $\lambda_i$ with incorrectly occupying the component at position $i$. The second term models interactions between sites where the correct (or incorrect) occupation of a single site determines the energy of another. The exact values of $\lambda_i$ and $\mu_{ij}$ could be chosen to ensure the ``correct" state ($\theta_i = \omega_i$ for all $i$) is non-degenerate as in the non-interacting model. The ellipses represent higher order interactions in this framework. Hamiltonians such as \refew{haminit_gen0} should be more physically relevant as they would correspond to systems where the energy cost for deviating from the lowest energy permutation is not simply linear but could be represented as a tensor valued fitting function.

We can make progress in studying the thermodynamics of more general Hamiltonians like \refew{haminit_gen0} by first only considering first- and second-order interaction terms and taking the interactions to be constants: $\lambda_i = \lambda_1$ for all $i$; $\mu_{ij} = \lambda_2/N$ for all $i,j$. The factor of $1/N$ is chosen so that the second term matches the extensive scaling of the first term. The partition function for such parameter selections is then
\begin{align}
Z_{N}(\beta;\lambda_1,\lambda_2 ) &  = \sum_{\vec{\theta} \,\in\, Sym(\omega)} \text{exp}\left(-\beta \lambda_1 \sum_{i=1}^N I_{\theta_{i} \neq \omega_{i}}-\right.\mm
& \qquad \left. \frac{\beta\lambda_2}{2N}\sum_{i, j =1}^N I_{\theta_i \neq \omega_i} I_{\theta_j \neq \omega_j} \right),
\end{align}
where $\lambda_1$ and $\lambda_2$ are interaction parameters with units of energy. We can also write this partition function in the \refew{jdef} basis as 
\begin{equation}
Z_{N}(\beta;\lambda_1, \lambda_2) = \sum_{j=0}^{N} g_{N}(j) e^{-\beta {\cal E}(j)  },
\end{equation}
where $g_{N}(j)$ is defined in \refew{gNj} and 
\begin{equation}
{\cal E}(j) = \lambda_1 j +\frac{\lambda_2}{2N} j^2
\label{eq:energy}
\end{equation}
is the energy function for the system.

\subsection{Calculating Order Parameter}

Our goal is to analyze the ``quasi"-phase behavior of this system in a way analogous to our analysis for the non-interacting system. To do so we begin with the Landau free energy function 
\begin{equation}
F_{N}(j, \beta) =  \lambda_1 j +\frac{\lambda_2}{2N}  j^2 -\frac{1}{\beta} \ln g_{N}(j).
\label{eq:free_en4}
\end{equation}
Alternative starting points for this derivation are presented in Appendix \ref{app:alt_deriv}. Our system is constitutively discrete, so it is not precisely correct to discuss our free energy in the language of analysis, but given our expression for \refew{gNj} we can map this system to a continuous one which bears the same thermodynamic properties and for which analysis is appropriate. Specifically, if we take $j$ to be continuous and use the identity $\Gamma(x+1) = x!$, we can write 
\begin{equation}
g_{N}(j) =  \frac{\Gamma(N+1)}{\Gamma(j+1) \Gamma(N-j+1)}  \frac{\Gamma(j+1, -1)}{e},
\label{eq:gNjsub}
\end{equation}
With the approximation $\Gamma(j+1, -1)\simeq \Gamma(j+1)$ and the substitution \refew{gNjsub}, \refew{free_en4} then becomes
\begin{equation}
f_{N}(j, \beta) =  \lambda_1 j +\frac{\lambda_2}{2N}  j^2 +\frac{1}{\beta} \ln \Gamma(N-j+1) + f_0
\label{eq:free_en}
\end{equation}
where we defined our approximated free energy as $f_{N}(j, \beta)$ and collected the $j$ independent constants into $f_0$. Now \refew{free_en} is fully continuous and amenable to analysis. To find the thermodynamic equilibrium of this system, we need to find the value of $j$ for which $\partial f_{N}(j, \beta)/\partial j = 0$ and $\partial^2f_{N}(j,\beta)/\partial j^2 >0$. For the first condition we have 
\begin{equation}
\frac{\partial}{\partial j} f_{N}(j, \beta) =  \lambda_1  +\frac{\lambda_2}{N}  j -\frac{1}{\beta}\psi_{0}(N-j+1) = 0. 
\label{eq:fprime}
\end{equation}
For $x \geq 0.6$ we have 
\begin{equation}
\psi_0(x) \simeq \ln (x-1/2),
\label{eq:psi0approx}
\end{equation}
as can be affirmed by Taylor expansion or plots of each side. Since the argument of $\psi_0(N-j+1)$ is bounded below by 1, the approximation in \refew{psi0approx} can be applied to \refew{fprime}. With this substitution, and setting the result to be valid for the equilibrium value $j = \overline{j}$, we then find the constraint
\begin{equation}
e^{\beta\lambda_2\, \overline{j}/N} = -e^{-\beta \lambda_1}\left(\overline{j} - N-1/2 \right),
\label{eq:jcond}
\end{equation}
which has the solution 
\begin{equation}
\frac{\overline{j}}{N} = 1 - \frac{1}{\beta\lambda_2} W\left(\frac{\beta\lambda_2}{N} e^{\beta\lambda_1+ \beta\lambda_2}\right)+ {\cal O}\left(\frac{1}{N}\right),
\label{eq:jsoln}
\end{equation}
where $W$ is the (branch unspecified) Lambert W function, defined by 
\begin{align}
W(x e^{x}) = x.
\label{eq:W}
\end{align} 
To specify the branch of the $W$ which corresponds to a stable equilibrium we compute the second derivative of our free energy at this derived critical point. Doing so yields 
\begin{align}
\frac{\partial^2}{\partial j^2} f_{N}(j = \overline{j}, \beta) & \simeq  \frac{1}{N}\left(\lambda_2 +\frac{1}{\beta}\frac{1}{1-\overline{j}/N} \right) \mm
& = \frac{\lambda_2}{N}\left(1+ \frac{1}{W\left(\frac{\beta\lambda_2}{N} e^{\beta\lambda_1+ \beta\lambda_2}\right)} \right).
\end{align}
Thus \refew{jsoln} (for $\lambda_2>0$) yields a free energy minimum for 
\begin{equation}
W\left(\frac{\beta\lambda_2}{N} e^{\beta \lambda_1+ \beta \lambda_2}\right) > -1,
\label{eq:mincond}
\end{equation}
and yields a maximum for the inverse condition. This amounts to stating that the stable equilibrium for $\overline{j}$ is defined by the principal branch of the Lambert W function where $W = W_0 \geq -1$, and the unstable equilibrium for $\overline{j}$ is defined by the negative branch where $W = W_{-1} < -1$. 

Thus, the order parameter for this system is 
\begin{equation}
\frac{\overline{j}_0}{N} = 1 - \frac{1}{\beta\lambda_2} W_0\left(\frac{\beta\lambda_2}{N} e^{\beta\lambda_1+ \beta\lambda_2}\right)+ {\cal O}\left(\frac{1}{N}\right).
\label{eq:j0soln}
\end{equation}
We note that taking $\lambda_2 \to 0$ and using $W(x) = x + {\cal O}(x^2)$ for $|x|\ll 1$ returns us to the non-interacting result \refew{eos}. 

For completeness, we define the value of $j$ which yields a free energy maximum as $\overline{j}_{-1}$; it is related to \refew{j0soln} by replacing the principal branch function $W_0$ with $W_{-1}$.

\subsection{Discussion of Parameter Space}

In the previous section, we found that the order parameter for this system was given by \refew{j0soln}.
We noted that this solution represented a local minimum of the free energy as long as the Lambert W function satisfied $W = W_0 > -1$. Thus when this condition is violated, $\overline{j}_0$ is no longer a valid stable equilibrium, and our system has undergone a ``quasi"-phase transition or simply a transition.

Moreover, our values of $j$ are bounded below by $j=0$ and bounded above by $j= N$, neither conditions of which are naturally constrained by \refew{j0soln}. Thus these two conditions are associated with two other transitions. In all, then, there are three conditions which define the quasi-phase boundaries in this system. 

\begin{figure}[t]
\centering
\includegraphics[width= \linewidth]{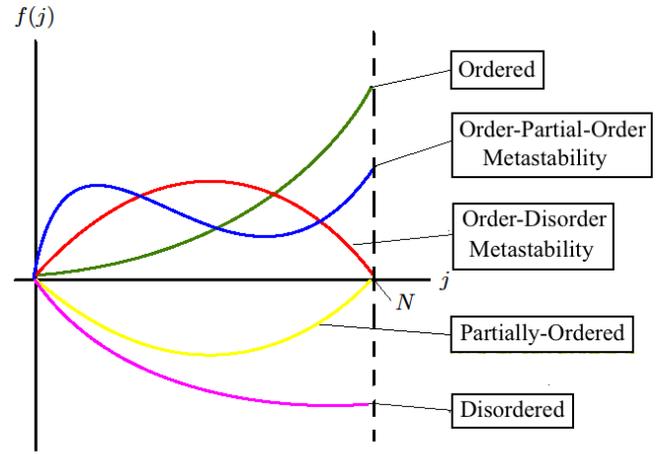}
	\caption{(Color online) Possible functional forms of \refew{free_en}: We note that the stabilities that define the $j=0$ and $j=N$ points are not thermodynamic stabilities (namely they don't arise from the $f'(j) = 0$ condition). Rather since the spectrum of $j$ values is bounded below by $0$ and above by $N$, owing to these boundary conditions the system can become trapped in ordinarily unstable parts of the free energy curve. The colors match the color of the associated region of parameter space in \reffig{phase_diagram5}. }
\label{fig:free_en}
\end{figure}

While there are three conditions which define transitions in this system, there are in fact five distinct regimes of parameter space.  We can obtain a qualitative sense of these regimes by creating schematic plots of the free energy \refew{free_en} for various parameter values of $\lambda_1$ and $\lambda_2$. The possible plots can be placed into five categories according to the plot's stable or metastable $j$ values. We depict these possible plots in \reffig{free_en}. We note that only the free energy plots with valid values of $\overline{j}_0$ contain what we normally consider a thermodynamic equilibrium; the other plots have ``stable" values of $j$ arising only from the $j=0$ and/or $j=N$ boundary conditions. 

Qualitatively, we can name the states according to the sequence space to which their equilibrium values of $j$ correspond. We know for $j=0$, our system is in a state with zero incorrect components in $\vec{\theta}$ and hence the system is ``perfectly ordered" or just "ordered". Conversely for $j=N$ our system has $N$ incorrect components and hence the system is ``completely disordered" or just ``disordered". The in-between case of $j = \overline{j}$ where $0 < \overline{j} < N$ can be given the related label of ``partially-ordered". Thus, the regime names associated with our possible values of the order parameters are 
\begin{itemize}
\item \tbf{Ordered Regime} {\bf ($\boldsymbol{j= 0}$ stable)}: Neither $\overline{j}_{0}$ or $\overline{j}_{-1}$ exist; $f(N, \beta)>0$.
\item \tbf{Disordered Regime} {\bf ($\boldsymbol{j = N}$ stable) }:  Neither $\overline{j}_{0}$ or $\overline{j}_{-1}$ exist; $f(N, \beta)<0$.
\item \tbf{Partially-Ordered Regime} {\bf ($\boldsymbol{j = \overline{j}_0}$ stable)}: Only $\overline{j}_0$ exists. 
\item \tbf{Order and Disorder Metastable Regime} {\bf ($\boldsymbol{j=0}$ and $\boldsymbol{j=N}$ stable)}: Only $\overline{j}_{-1}$ exists.
\item \tbf{Order and Partial-Order Metastable Regime} {\bf ($\boldsymbol{j = 0}$ and $\boldsymbol{j= \overline{j}_0}$ stable) }: Both $\overline{j}_0$ and $\overline{j}_{-1}$ exist.
\end{itemize}
We note that it seems to be a fundamental feature (or a lack of one) of this system, that the free energy \refew{free_en} does not admit a metastability between partial-order and disorder. 

\subsection{Monte-Carlo Generated Parameter Space}

\begin{figure*}[t]
\centering
\includegraphics[width= .75\linewidth]{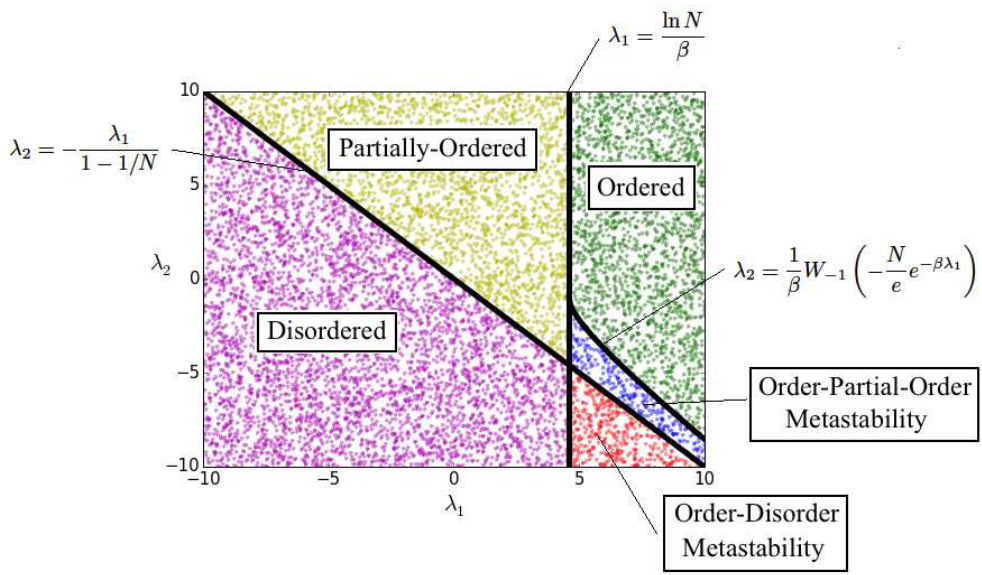}
	\caption{(Color online) $\lambda_1-\lambda_2$ Parameter Space for Interacting Mean Field System: We set $\beta = 1$ and $N=100$. We followed the Monte Carlo procedure outlined in the notes for 10,000 points. In the figure we also denoted the analytic lines (i.e., \refew{bc1}, \refew{bc2}, and \refew{bc3}) which define the separation between the phases. The colors correspond to the colors of the free energy curve in \reffig{free_en}.}
\label{fig:phase_diagram5}
\end{figure*}

With these regime definitions, we can depict the parameter space graphically. In \reffig{free_en} we showed the possible forms of the free energy for this system where each was categorized according to the existence of the local minimum critical point $\overline{j}_0$, the existence of the local maximum critical point $\overline{j}_{-1}$, and the sign of the quantity $f_{N}(j, \beta)$. We can extrapolate this categorization to $\lambda_1-\lambda_2$ parameter space, by determining which regions of parameter space correspond to specific plots in \refew{free_en}. Doing so through the Monte Carlo procedure described in Appendix \ref{app:mcpro}, we generated 10,000 points of the parameter space diagram in \reffig{phase_diagram5} for $\beta$ set to 1. We note that the parameter space exhibits five regimes separated by three lines cited in Table \ref{tab:table1} each of which correspond to the three conditions mentioned at the beginning of this section. These lines can be derived analytically (as shown in Appendix \ref{app:phase_bound}) by considering the conditions in turn and which regimes they serve to connect. 

\begin{table}[t]
\caption{\label{tab:table1}%
Functions defining boundaries between parameter regimes. 
}
\begin{ruledtabular}
\begin{tabular}{p{3.5cm} l}
\textrm{Regime Transition} & \textrm{Boundary in Parameter Space}  \\ 
 \hline Order/Partial-Order  & \vspace{.2cm} $\displaystyle \lambda_1 = \frac{1}{\beta}\ln N$ \\
Order/Order-Partial-Order Metastability & $\displaystyle \lambda_2 = \frac{1}{\beta} W_{-1}\left(- \frac{N}{e} e^{-\beta\lambda_1}\right)$\\
Order/Disorder & $\displaystyle \lambda_2 = - \frac{\lambda_1}{1-1/N}$ \\
\end{tabular}
\end{ruledtabular}
\end{table}

\subsection{Triple and Quadruple Points and Transition Temperatures}

From \reffig{phase_diagram5} we see that our system is characterized by two points where there is a coexistence between multiple regimes. Given that $W_{-1}(-e^{-1}) = -1$, we have that the Ordered, Partially-Ordered, and Order-Partially-Ordered Metastability coexistence point is characterized by the condition 
\begin{equation}
\lambda_1 = \ln N/\beta \quad \text{and} \quad \lambda_2 = -1/\beta.
\end{equation}
These conditions characterize the system's triple point. 

Similarly, for $N\gg1$, the Partially-Ordered, Disordered, Order-Partially-Ordered Metastability, and Order-Disorder Metastability coexistence point is characterized by the condition 
\begin{equation}
\lambda_1  =  \ln N/\beta \simeq - \lambda_2.
\end{equation}
This condition characterizes the quadruple point of the system. 

%

\reffig{phase_diagram5} also depicts the possible regimes of our system for a given temperature and various Hamiltonian parameters $\lambda_1$ and $\lambda_2$. More physically, we may be interested in knowing what are the ``quasi" phase properties of a system with a fixed $\lambda_1$ and $\lambda_2$ and a variable temperature. That is, what are the temperatures which define the various transitions between regimes in the system?

An arbitrary permutation system for a fixed $N$ and at a variable temperature is characterized by a specific energy function \refew{energy}. Such a system is therefore defined by a specific $\lambda_1$ and $\lambda_2$, and the system can be associated with a particular point (and hence region) in the parameter space of \reffig{phase_diagram5}. As we vary the temperature of this system, the temperature dependent regime-coexistence lines change and if they change in such a way as to extend the region of a regime to newly encompass our original point then our system has undergone a transition. In this way, we can define the temperatures which characterize various possible transitions of this system.

First, from \reffig{phase_diagram5} and \refew{bc3} we note the regime-coexistence line between the partially-ordered and disordered regime is independent of temperature, and so there is no critical temperature defining a partial-order to disorder transition.

From \refew{bc1}, we can infer that the partial-order to order transition is characterized by moving below the temperature
\begin{equation}
{k_B T_{c1}} = \frac{\lambda_1}{\ln N}. 
\end{equation}
Contingent on which region of parameter space the system lies, this temperature also characterizes the disorder to order-disorder metastability transition and the partial-order to order-partial-order metastability transition. 

And from \refew{trans2}, we can solve for the associated transition temperature given fixed $\lambda_1$ and $\lambda_2$ to find
\begin{equation}
{k_B T_{c2}} = {(\lambda_1 + \lambda_2)}\left[W_0\left(- \frac{N}{e\lambda_2}(\lambda_1+ \lambda_2) \right)\right]^{-1}
\end{equation}
where this expression is only relevant for $-\lambda_1<\lambda_2<0$ and $\lambda_1> k_{B} T\ln N $. Moving above this temperature leads to the order to order-partial-order metastability transition.

\section{Discussion \label{sec:five}}

In this work, motivated by an abstraction of a foundational problem in protein design, we posited and analyzed the basic properties of a statistical physics model of permutations. Formally, we considered a simple statistical physics model where the space of states for $N$ lattice sites was isomorphic to the symmetric group of degree $N$ \cite{dixon1996permutation}, and where the energy of each permutation was a function of how much the permutation deviates from the identity permutation. 

In this model, we found that due to a state space which could not be factorized in a basis defined by lattice sites, even the superficially non-interacting system can exhibit phase-like transitions, i.e., temperature dependent changes in the value of the order parameter which do note exhibit the properties typically associated which phase transitions in infinite systems. When interactions are introduced through a quadratic mean field term, the system is capable of exhibiting five regimes of thermal behavior, and is characterized by two transition-temperatures corresponding to various quasi-phase transitions. 

The introduced model provides us with a basic exactly soluble system for certain interaction assumptions and thus provides a concrete model-based understanding of a system with a non-factorizable state space. Because of its utility and the type of results obtained, the model deserves to be subject to the standard extensions of typical canonical models in statistical mechanics. In particular we hope to extend it to non-trivial site dependent interactions. For example, a nearest neighbor interaction Hamiltonian of the kind which characterize the Ising Model,
\begin{equation}
{\cal H}(\{\theta_i \}) = - q \sum_{i=1}^N I_{\theta_i \neq \omega_i} I_{\theta_{i+1} \neq \omega_{i+1}},
\end{equation}
would be an alternative physical extreme to the mean-field interactions considered in Section \ref{sec:three}. 

We could also consider a generalized chain of components where the interactions between sites or the cost for an incorrectly filled site is not constant but is drawn from a distribution of values. Such a system of quenched disorder would characterize a permutation glass which may contain interesting results due to the unique nature of the state space. 

Supposing it is possible to define more interesting interactions models, a natural investigation would concern the renormalization group properties of the system. Specifically, we would be interested in how would one sum over specific states (as characteristic of a renormalization group transformation) when the state space of a system looks like, 
\begin{equation}
{\cal S}_{\text{system}} = \prod_{i=1}^{N} \otimes \, {\cal S}_{i}
\end{equation}
i.e., is not factorizable along lattice sites. 

Finally, to connect this model of permutations to problems more relevant to protein design it would prove necessary to incorporate the possibility of repeated components or the background geometry of a lattice chain.

\begin{acknowledgments}
MW thanks Verena Kaynig-Fittkau for her unpublished work that inspired this investigation and Amy Gilson, Abigail Plummer, Vinothan Manoharan, and Michael Brenner for helpful discussions. 
\end{acknowledgments}

\appendix

\section{Alternative Derivations of {\refew{jcond}} \label{app:alt_deriv}}

\subsection{Hubbard-Stratonovich Approach}
%
We re-derive \refew{jcond} using the Hubbard-Stratonovich method. We start this derivation assuming $\lambda_2 = -|\lambda_2|$; we will later see our resulting free energy can be analytically continued to the $\lambda_2 >0$ case. 

For $\lambda_2 = - |\lambda_2|$, the partition function is
\begin{equation}
Z_{N}(\beta; \lambda_1,\lambda_2) = \sum_{j=0}^{N}g_{N}(j) e^{-\beta\lambda_1 j + \beta|\lambda_2|j^2/2N}. 
\end{equation}
Then, applying the identity
\begin{equation}
e^{\beta|\lambda_2|j^2/2N} = \sqrt{\frac{N}{2\pi\beta |\lambda_2|}}\int^{\infty}_{-\infty} dx \, e^{-Nx^2/2\beta|\lambda_2|- j x},
\end{equation}
we have
\begin{widetext}
\begin{align}
Z_{N}(\beta; \lambda_1,\lambda_2)  & = \sqrt{\frac{N}{2\pi\beta |\lambda_2|}}\int^{\infty}_{-\infty} dx \, e^{-Nx^2/2\beta|\lambda_2|} \sum^{N}_{j =0} g_{N}(j) e^{-j(\beta \lambda_1+x)}\mm
& = \sqrt{\frac{N}{2\pi\beta |\lambda_2|}}\int^{\infty}_{-\infty} dx \, e^{-Nx^2/2\beta|\lambda_2|} Z_{N}(\beta\lambda_1+x),
\label{eq:hbbd}
\end{align}
where $Z_{N}(x) \equiv \sum_{j=0}^{N} g_{N}(j) e^{-j x}$. From \refew{partfunc} we found 
\begin{equation}
Z_{N}(x) = \int^{\infty}_{0} ds\, e^{-s} \left(1+ (s-1)e^{-x}\right)^N.
\end{equation}
So \refew{hbbd} becomes 
\begin{equation}
Z_{N}(\beta; \lambda_1,\lambda_2)  =  \sqrt{\frac{N}{2\pi\beta |\lambda_2|}}\int^{\infty}_{0}ds\,\int^{\infty}_{-\infty} dx \,  \left(1+ (s-1)e^{-\beta\lambda_1-x}\right)^{N}e^{-s - Nx^2/2\beta|\lambda_2|}.
\label{eq:hbbd2}
\end{equation}
\end{widetext}
The function to which we apply steepest descent is then 
\begin{equation}
h(s,x) = s +\frac{Nx^2}{2\beta|\lambda_2|} - N\ln  \left(1+ (s-1)e^{-\beta\lambda_1-x}\right).
\end{equation}
Computing the conditions for $\partial_{s}h(s= \overline{s},x = \overline{x}) = 0$ and $\partial_x h(s= \overline{s},x = \overline{x}) = 0$ we obtain, respectively, 
\begin{align}
1 - N\frac{e^{-\beta\lambda_1 - \overline{x}}}{1+ (\overline{s}-1)e^{-\beta\lambda_1-\overline{x} }} & = 0\\
\frac{\overline{x}}{\beta|\lambda_2|} +  \frac{  (\overline{s}-1)e^{-\beta\lambda_1-\overline{x} } }{ 1+ (\overline{s}-1)e^{-\beta\lambda_1-\overline{x} } } &= 0.
\end{align}
Solving for $\overline{s}$ in the first equation we have
\begin{equation}
\overline{s} = N + 1 - e^{\beta\lambda_1 + \overline{x}}
\end{equation}
and with the second equation we have the condition 
\begin{equation}
\frac{\overline{x}}{\beta|\lambda_2|} = -\frac{1}{N} (\overline{s}-1).
\end{equation}
Substituting the second condition into the first yields
\begin{equation}
\overline{s}-1 = N  - e^{\beta\lambda_1 -{\beta|\lambda_2|}(\overline{s}-1)/N },
\end{equation}
or 
\begin{equation}
e^{-{\beta|\lambda_2|}(\overline{s}-1)/N } = e^{\beta\lambda_1}\left( N - (\overline{s}-1)\right). 
\end{equation}
Thus, the solution for $\overline{s}$ can be expressed in terms of the Lambert W function as 
\begin{equation}
\frac{\overline{s} -1}{N} = 1 + \frac{1}{\beta|\lambda_2|} W\left(-\frac{\beta |\lambda_2|}{N} e^{\beta\lambda_1 -{\beta|\lambda_2|}}\right).
\label{eq:sol}
\end{equation}
For $\lambda_2>0$, we can employ the complex version of the Hubbard-Stratonovich identity:
\begin{equation}
e^{-\beta\lambda_2j^2/2N} = \sqrt{\frac{N}{2\pi\beta \lambda_2}}\int^{\infty}_{-\infty} dx \, e^{-Nx^2/2\beta\lambda_2- ijx}.
\end{equation}
Working through an analogous steepest descent procedure, we find that the equilibrium value for $\overline{s}$ is 
\begin{equation}
\frac{\overline{s} -1}{N} = 1- \frac{1}{\beta\lambda_2} W\left(\frac{\beta \lambda_2}{N} e^{\beta\lambda_1+{\beta\lambda_2}}\right) \equiv \frac{\overline{j}}{N},
\end{equation}
which could have been extrapolated from \refew{sol} by taking $|\lambda_2| \to - \lambda_2$. From this expression for the equilibrium condition, and as an analogy with the non-interacting case, it turns out the order parameter in this case is not $\overline{s}$ but rather $\overline{s}-1$. 

\subsection{Gibbs-Bogoliubov Inequality Derivation of \refew{jcond}}
We re-derive \refew{jcond} using the Gibbs-Bogoliubov Inequality\cite{yeomans1992statistical}. The inequality is
\begin{equation}
F[{\cal H}] \leq F[{\cal H}_0 ] + \langle {\cal H} - {\cal H}_0\rangle_0.
\label{eq:gbineq}
\end{equation}
The Hamiltonian which defines our system is 
\begin{equation}
{\cal H} = \lambda_1 \sum^{N}_{i=1} I_{\theta_i \neq \omega_i} + \frac{\lambda_2}{2N} \sum_{i, j} I_{\theta_i \neq \omega_i} I_{\theta_j \neq \omega_j} \equiv \lambda_1 j + \frac{\lambda_2}{2N} j^2,
\end{equation}
and our variational Hamiltonian is instead
\begin{equation}
{\cal H}_0 = \lambda_0 \sum^{N}_{i=1} I_{\theta_i \neq \omega_i} = \lambda_0 j.
\label{eq:var_ham}
\end{equation}
From \refew{partfunc} we know 
\begin{align}
F[{\cal H}_0] & = - \frac{1}{\beta} \ln Z_{N}(\beta\lambda_0)\mm
& = - \frac{1}{\beta} \ln \left\{ \int^{\infty}_{0} ds\, e^{-s} \left(1+(s-1)e^{-\beta\lambda_0}\right)^N\right\}.
\end{align}
We can also define 
\begin{equation}
\langle {\cal O}(j) \rangle_{0, N} = \sum_{j=0}^{N} {\cal O}(j)\, e^{-\beta\lambda_0 j},
\end{equation}
as the average with respect to our variational Hamiltonian \refew{var_ham}. 
Thus, \refew{gbineq} becomes 
\begin{align}
F[{\cal H}]  &\leq - \frac{1}{\beta} \ln Z_{N}(\beta\lambda_0)  + (\lambda_1 - \lambda_0) \langle j \rangle_{0,N} +\frac{\lambda_2}{2N}\langle j^2 \rangle_{0,N}\mm
&  \equiv f(\lambda_0)
\end{align}
Differentiating $f$ with respect to $\lambda_0$ allows us to compute the maximum of this quantity. Given $\langle j \rangle_{0,\,N} = - \partial \ln Z_{N}(\beta\lambda_0)/\partial(\beta\lambda_0)$, we then find
\begin{equation}
f'(\lambda_0)  = (\lambda_1 - \lambda_0) \frac{\partial }{\partial \lambda_0}\langle j \rangle_{0,N} + \frac{\lambda_2}{2N}\frac{\partial }{\partial \lambda_0}\langle j^2 \rangle_{0,N},
\end{equation}
which if we take to be zero at some $\lambda_0 = \lbar$ gives us 
\begin{equation}
 0 = (\lambda_1 - \lbar) \frac{\partial }{\partial \lambda_0}\langle j \rangle_{0,N}\Big|_{\lambda_0 = \lbar} + \frac{\lambda_2}{2N}\frac{\partial }{\partial \lambda_0}\langle j^2 \rangle_{0,N}\Big|_{\lambda_0 = \lbar}.
 \label{eq:lamb_eq}
\end{equation}
To compute these derivatives we make use of various identities. First we note
\begin{equation}
\langle j^2 \rangle_{0,N} = \frac{1}{Z_{N}(\beta\lambda_0)} \frac{\partial^2}{\partial(\beta\lambda_0)^2} Z_{N}(\beta\lambda_0),
\end{equation}
so 
\begin{align}
\frac{\partial}{\partial (\beta\lambda_0)}\langle j \rangle_{0,N} & = -\frac{\partial^2}{\partial (\beta\lambda_0)^2}\ln Z_{N}(\beta\lambda_0) \mm
& = - \frac{1}{Z_{N}(\beta\lambda_0)} \frac{\partial^2}{\partial(\beta\lambda_0)^2} Z_{N}(\beta\lambda_0)\mm
& \quad + \frac{1}{Z_{N}(\beta\lambda_0)^2} \left(\frac{\partial}{\partial(\beta\lambda_0)} Z_{N}(\beta\lambda_0)\right)^2\mm
& = - \langle j^2 \rangle_{0,\,N} + \langle j \rangle_{0,\,N}^2. 
\end{align}
This last equality implies 
\begin{equation}
\frac{\partial}{\partial (\beta\lambda_0)}\langle j^2 \rangle_{0,N}  = -\frac{\partial^2\langle j \rangle_{0,N}}{\partial (\beta\lambda_0)^2} + 2\langle j \rangle_{0,N}\frac{\partial \langle j \rangle_{0,N} }{\partial (\beta\lambda_0)}, 
\end{equation}
and so \refew{lamb_eq} becomes 
\begin{align}
 0 & = \left[\lambda_1 - \lbar + \frac{\lambda_2}{N}\langle j \rangle_{0,N} \right] \frac{\partial }{\partial \lambda_0}\langle j \rangle_{0,N}\Big|_{\lambda_0 = \lbar}\mm
 & \quad  - \frac{\lambda_2}{2N\beta}\frac{\partial^2\langle j \rangle_{0,N}}{\partial \lambda_0^2}\Big|_{\lambda_0 = \lbar}
  \label{eq:lamb_eq2}
\end{align}
To compute these quantities we need to approximate the partition function for our variational system. Using the method of steepest descent
\begin{align}
Z_{N}(\beta\lambda_0) &=  \int^{\infty}_{0} ds\, e^{-s} \left(1+(s-1)e^{-\beta\lambda_0}\right)^N \mm
& =  \sqrt{2\pi N} \left(\frac{N}{e^{\beta\lambda_0}}\right)^N \exp\left(e^{\beta\lambda_0} - N-1\right)\mm
& \quad \times \left(1 + {\cal O}\left(N^{-1}\right)\right),
\end{align}
and so we have 
\begin{align}
\langle j  \rangle_{0, N} & = - \frac{\partial}{\partial(\beta\lambda_0)} \ln Z_{N}(\beta\lambda_0) \mm
& = N - e^{\beta\lambda_0} +{\cal O}\left(N^{-1}\right).
\label{eq:jgibbs}
\end{align}
Computing the relevant derivatives in \refew{lamb_eq2} we find
\begin{align}
 0 & = \left[\lambda_1 - \lbar + \frac{\lambda_2}{N}\langle j \rangle_{0,N} \right] \left(- e^{\beta\lambda_0} +{\cal O}\left(N^{-1}\right)\right) \mm
& \hspace{1cm} - \frac{\lambda_2}{2N}\left(- e^{\beta\lambda_0} +{\cal O}\left(N^{-1}\right)\right) = 0 \mm
& = \left[\lambda_1 - \lbar + \frac{\lambda_2}{N}\langle j \rangle_{0,N}- \frac{\lambda_2}{2N}\right]e^{\beta\lambda_0} +{\cal O}\left(N^{-1}\right) 
\end{align}
Neglecting subleading terms of ${\cal O}(1/N)$ (a choice only valid for $\langle j \rangle_{0,N} \gg 1$), solving for $\lbar$, and using \refew{jgibbs} we then obtain the equilibrium constraint
\begin{equation}
e^{\beta \lambda_1 - \beta\lambda_2/2N + \beta\lambda_2\langle j \rangle_{0,N}/2N} = N - \langle j \rangle_{0,N},
\end{equation}
which when solved for $\langle j \rangle_{0,N}/N$ gives us
\begin{equation}
\langle j \rangle_{0,N}/N = 1 - \frac{1}{\beta\lambda_2} W\left(\frac{\beta\lambda_2}{N} e^{\beta\lambda_1 + \beta\lambda_2\left( 1- \frac{1}{2N}\right)} \right),
\end{equation}
or, given our approximations and limiting expressions, the result
\begin{equation}
\langle j \rangle_{0,N}/N = 1 - \frac{1}{\beta\lambda_2} W\left(\frac{\beta\lambda_2}{N} e^{\beta\lambda_1 + \beta\lambda_2} \right) + {\cal O}\left(N^{-1}\right).
\end{equation}

\section{Monte-Carlo Procedure for Parameter Space \label{app:mcpro}}
To generate \reffig{phase_diagram5}, we implemented the following MC algorithm:
\begin{enumerate}
\item Uniformly sample two points for $\lambda_1$ and $\lambda_2$ separately from within a certain bounded domain. 
\item Draw the free energy curve \refew{free_en} corresponding to the sampled values $(\lambda_1, \lambda_2)$.
\item Label the curve according to which schematic curve in \reffig{free_en} it corresponds (i.e., according to its $\overline{j}_0$, $\overline{j}_-$ and $f_N(N, \beta)$ properties).
\item Color the point to signify the label. 
\end{enumerate}
We repeated this procedure for $10,000$ points with $\beta =1$. The regime separation lines were included after the MC procedure from the analytic forms cited in the text.

\section{Analytic Functions of Regime Boundaries \label{app:phase_bound}}

\subsubsection{Order to Partial-Order Transition}
The regime boundary which separates the ordered and the partially ordered regime is defined by the condition $\overline{j}_0 \geq  0$. For this regime boundary we have the condition 

\begin{equation}
 1 - \frac{1}{\beta\lambda_2} W_0\left(\frac{\beta\lambda_2}{N} e^{\beta\lambda_1+ \beta\lambda_2}\right)\geq 0 ,
\end{equation}
or
\begin{equation}
W_0\left(\frac{e^{\beta\lambda_1}}{N} \beta\lambda_2 e^{\beta\lambda_2}\right)\geq \beta\lambda_2.
\end{equation}
From a plot of $W_0(a \,x e^{x})/x$ for real $a$, we see that $W(a x e^{x})/x >1$ if $a>1$ and $W(a x e^{x})/x<1$ if $a<1$. Thus this order to partial-order transition is defined by the condition $e^{\beta\lambda_1}/N = 1$, or 
\begin{equation}
\lambda_1 = \frac{\ln N}{\beta}.
\label{eq:bc1}
\end{equation}

\subsubsection{Order to Order-Partial-Order Metastability Transition}
The regime boundary which separates the ordered regime from the order and partial-order metastability regime is defined by the condition $- 1 \leq W < 0 $. This condition is where the $\overline{j}_0$ and $\overline{j}_-$ begin coexisting \cite{weisstein2002lambert}, thus creating the mutual existence of a local maxima and local minima in \reffig{free_en}. Thus for this regime boundary we have the condition 
\begin{equation}
- 1 \leq W\left(\frac{\beta\lambda_2}{N} e^{\beta\lambda_1+ \beta\lambda_2}\right)< 0,
\end{equation}
This condition is valid so long as the argument of $W$ satisfies
\begin{equation}
- e^{-1} \leq \frac{\beta\lambda_2}{N} e^{\beta\lambda_1+ \beta\lambda_2}< 0.
\end{equation}
This inequality can only possibly be satisfied for $\lambda_2 <0$ and if $\lambda_2<0$ the right inequality is automatically true. So, our transition condition is given by 
\begin{equation}
- \frac{N}{e} e^{-\beta\lambda_1} = \beta\lambda_2 e^{\beta\lambda_2}.
\label{eq:trans2}
\end{equation}
In the Order phase we automatically have $\beta\lambda_1 > \ln N$, so the LHS of \refew{trans2} is greater than or equal to $-e^{-1}$. Moreover, since $\lambda_2$ is exclusively negative, at $\beta\lambda_2 = \ln N$,  $\beta\lambda_1$ is at a maximum value of $\beta\lambda_2 = -1$. For $\beta\lambda_2 \leq -1$, the solution to \refew{trans2} is then
\begin{equation}
\lambda_2 = \frac{1}{\beta} W_{-1}\left(- \frac{N}{e} e^{-\beta\lambda_1}\right).
\label{eq:bc2}
\end{equation}

\subsubsection{Order to Disorder Transition}
The regime boundary which separates the partially ordered regime from the disordered regime is defined by the condition $\overline{j}_0 \leq N-1$. We set the maximum value of $\overline{j}_0$ to $N-1$ rather than $N$ because \refew{j0soln} is associated with a free energy which diverges at $\overline{j}_0 = N$ and this approximate result is thus only physical up to $N-1$. Alternatively we could see the maximum condition $\overline{j}_0 = N-1$ as respecting the fact that \refew{j0soln} is only valid up to ${\cal O}(N^{-1})$. For this regime boundary we have the condition
\begin{equation}
1- 1/N  \geq 1 - \frac{1}{\beta\lambda_2} W_0\left(\frac{\beta\lambda_2}{N} e^{\beta\lambda_1+ \beta\lambda_2}\right)
\end{equation}
or 
\begin{equation}
\frac{\beta\lambda_2}{N}  \leq W_0\left(\frac{\beta\lambda_2}{N} e^{ \beta\lambda_2/N} e^{\beta\lambda_1- \beta\lambda_2 - \beta\lambda_2/N}\right).
\end{equation}
Again, using the condition that $W_0(a\, xe^{x})/x > 1$ if $a>1$, we find that the critical condition for this transition is $\beta\lambda_1+ \beta\lambda_2 - \beta\lambda_2/N=0$ or 
\begin{equation}
\lambda_2 = - \frac{\lambda_1}{1-1/N}.
\label{eq:bc3}
\end{equation}

\newcommand{\noopsort}[1]{} \newcommand{\printfirst}[2]{#1}
  \newcommand{\singleletter}[1]{#1} \newcommand{\switchargs}[2]{#2#1}
\end{document}